\documentclass{bioinfo}
\copyrightyear{2015}
\pubyear{2015}

\usepackage{url}

\DeclareGraphicsExtensions{.eps}

\newcommand{\reffig}[1]{Figure \ref{#1}}

\begin{document}
\firstpage{1}

\title[LeoTask]{LeoTask: a fast, flexible and reliable framework for computational research}
\author[Zhang \textit{et~al}]{Changwang Zhang\,$^{1,2,}$\footnote{to whom correspondence should be addressed}, Shi Zhou\,$^{1}$ and Benjamin M. Chain\,$^3$}
\address{$^{1}$Department of Computer Science, $^2$Security Science Doctoral Research Training Centre, $^3$Division of Infection and Immunity, University College London, UK}

\history{}

\editor{}

\maketitle

\begin{abstract}

\section{Summary:} LeoTask is a Java library for computation-intensive and time-consuming research tasks. It automatically executes tasks in parallel on multiple CPU cores on a computing facility. It uses a configuration file to enable automatic exploration of parameter space and flexible aggregation of results, and therefore allows researchers to focus on programming the key logic of a computing task. It also supports reliable recovery from interruptions, dynamic and cloneable networks, and integration with the plotting software Gnuplot.

\section{Availability and implementation:}
The source code for LeoTask is freely available under FreeBSD License at \url{https://github.com/mleoking/leotask}. 

\section{Contact:} \href{changwang.zhang.10@ucl.ac.uk}{changwang.zhang.10@ucl.ac.uk}

\end{abstract}

Research tasks, especially in the field of bioinformatics, are increasingly computationally intensive \cite{Zomaya_2006}. Computational research (e.g. simulations and data analysis) typically explores results over a large parameter space and often needs to repeat a task a number of times to get an average result. Many complex computational tasks would last for days, or even weeks. As a result, they are prone to artificial (e.g. a colleague sharing the same computing facility stops your program) or natural (e.g. power outage) interruptions.

To accelerate the research, it is imperative to conduct tasks in parallel, fully utilising the processing power of computing facilities \cite{Zomaya_2006}. Nowadays, computing facilities normally have multiple cores in their Central Processing Unit (CPU), and each of the cores can individually conduct a processing task \cite{Geer_2005, Blake_2009}. For example, a latest desktop computer can have 4 to 8 cores, while a computing sever can have more than 16 cores.

While there are built-in mechanisms for parallel task running in all major programming languages, the level of complexity often makes these built-in mechanisms difficult to use and time consuming to program with. For example, the built-in parallel running mechanism in Java requires choosing the parts of the program that can be accessed by multiple tasks in parallel and the parts of the program that should be accessed by only one task at a time. Such choices affect not only the speed of a program but also its accuracy, i.e. a wrong choice could end up with a slow program and a program that gives wrong results.

Reliability is a critical but often overlooked feature of many computational programs and frameworks. A reliable program should be able to recover and continue running from interruptions. Much effort is often needed to make a program reliable, especially if the program runs in parallel. 

\begin{figure}[h]\centering
\includegraphics[width=0.3\textwidth]{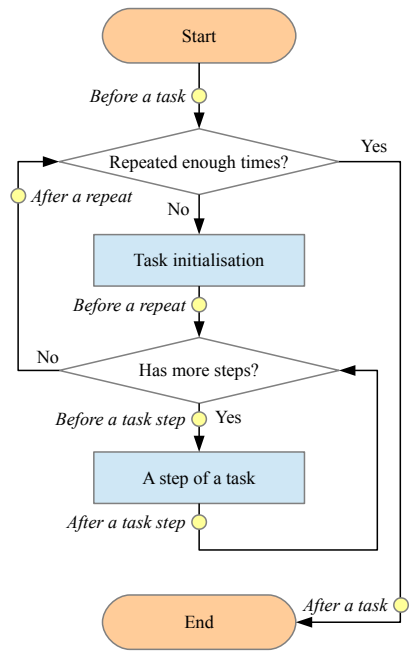}
\caption{\label{fig-flow} Default flow and time points for a task. There are 6 default built-in time points for result collection: 1) before the start of a task, 2) before starting a repeated run of a task, 3) before starting a step in a run, 4) after finishing a step in a run, 5) after finishing a repeated run of a task, 6) after finishing a task.}
\end{figure}

Here we present a framework, LeoTask, to facilitate reliably conducting research tasks in parallel. It has the following combination of features which would be attractive to the research community:

\begin{itemize}
\item \textit{Automatic \& parallel parameter space exploration}

LeoTask uses an intuitive configuration file to specify the value ranges of task parameters. The framework will figure out all possible combinations of values for all parameters and then run tasks with different parameter value combinations in parallel, i.e. LeoTask automatically explores the parameter space. The framework maps and runs all the tasks to available processing cores of a computing facility. 

\item \textit{Flexible \& configuration-based result aggregation}

The configuration file also sets when and how task results are aggregated. As shown in \reffig{fig-flow}, LeoTask has a default task flow with 6 default time points for specifying when the results are collected. Applications using the framework can also use different task flows and define additional time points. The framework supports aggregating results conditioned on a set of parameters. For example, for a task with parameter $x_1$, $x_2$ and result $y$. Given the value range of $x_1$ and $x_2$, the framework can aggregate $y$ conditioned on the value of $x_1$, $x_2$, value pair ($x_1$, $x_2$), or a any mathematical function of $x_1$ and $x_2$.

\item \textit{Programming model focusing only on the key logic}

The framework separates the key logic of a task from other ``bookkeeping'' parts of the program, which include parameter setting, result aggregation, result plotting etc. Users only need to program the key logic and use the configuration file to set other ``bookkeeping'' parts of a task. This facilitates sharing a program among a community where many end users only need to rerun a program with different parameter values and different ways to aggregate results. The framework frees those end users from reading the program and they only need to change a more intuitive configuration file.

\item \textit{Reliable \& automatic interruption recovery.}

LeoTask periodically writes the current state and results of the tasks to checkpoint files and can then recover and continue running from a checkpoint file when necessary. To recover from an interruption and still guarantee the completeness and correctness of results, a program has to track the tasks running in parallel. The framework handles these technique details automatically.

Considering that Java programs can run \textit{directly} on all major computing facilities (with different operating systems), this feature provides not only just reliability but also flexibility for program running. The following scenario is possible for a program using LeoTask: a user started the program in a laptop first; after several hours he / she found that it was too slow, thus copied the program and a check point file to a server A and continued running the program there; after another several hours, the running speed decreased on server A because some other colleagues started to use server A and made it busy, so he / she copied the program and a check point file from sever A to another unused server B and finished the remaining part of the program there.

\item \textit{Dynamic \& cloneable networks structures.}

Network analysis is widely used in many research tasks especially in bioinformatic studies \cite{Boccaletti_2006, Miller_2012}. For example, there are transcription networks \cite{Roy_2008}, signalling networks \cite{Papin_2005}, epidemic spreading networks \cite{Newman_Book_2010}, etc. LeoTask provides data structures to represent a node, a link, a network, and a network set. A network consists of nodes and links. A network set includes multiple networks and they can overlap (share nodes) with each other. All structures can be dynamic \cite{Zschaler_2013}, supporting adding, deleting, and updating elements. The framework also allows merging multiple networks.

LeoTask can create copies of a large network system including the states of each node, link, network, and network set. This enables users to investigate many common scenarios that are previously time consuming to program. For example, a user can simulate an epidemic spreading for the first 10 minutes and make 5 copies of the current network system (preserving the states of networks, nodes, and links) to test 5 different intervention strategies on these 5 copies \textit{in parallel} so that the initial condition are random but at the same time \textit{exactly} the same for these different strategies tested.

\item \textit{Integration with Gnuplot.} 

Gnuplot is a piece of widely-used open-source plotting software. LeoTask can output aggregated results directly as Gnuplot scripts which can be processed by Gnuplot to produce publication-quality figures. In addition, LeoTask includes a unique hybrid programming engine that compiles Gnuplot scripts and enables users to refer to values of Java objects or call Java functions within Gnuplot scripts.

\end{itemize}

The LeoTask framework has been used in epidemic spreading simulations \cite{Zhang_hm_2014}, large dataset analysis \cite{Zhang_conficker_2014}, and HIV model parameter fitting from clinical records. An introduction illustrating how to use the framework in an example application is available from \url{https://github.com/mleoking/leotask/blob/master/leotask/introduction.pdf?raw=true}.

We believe LeoTask's combination of features makes it useful for a wide research community. We also welcome suggestions or directly contributions to improve LeoTask and make it more widely available. 

\bibliography{references}

\end{document}